\documentclass[aps,prl,floats,superscriptaddress,nofootinbib,floatfix,twocolumn,longbibliography,showkeys]{revtex4-1}
\usepackage{amssymb,amsmath,amsfonts,amsthm}
\usepackage{graphicx}
\usepackage{newtxtext,newtxmath}
\usepackage{dcolumn}
\usepackage{bm}
\usepackage{comment}
\usepackage{csquotes}
\usepackage{mathtools}
\usepackage{gensymb}
\usepackage{hyperref}
\usepackage{makecell}
\usepackage{changepage,orcidlink}
\usepackage{multirow}
\usepackage[table]{xcolor}
\usepackage[mathscr]{euscript}
\usepackage{ulem}
\usepackage{comment}
\graphicspath{{./fig/}}

\hypersetup{
	colorlinks=true,
	linkcolor=blue,
	filecolor=blue,      
	urlcolor=blue,
	citecolor=blue,
}

\begin{document}

\title{Re-entrance or monotonic: How chemotaxis controls pattern formation}
\author{Mintu Karmakar\,\orcidlink{0000-0003-1037-2580}}
\email{mkarmakar094@ucas.ac.cn, mkarmakar094@gmail.com}
\affiliation{Wenzhou Institute of the University of Chinese Academy of Sciences, Wenzhou, Zhejiang 325011, China.}
\affiliation{School of Physical Sciences, University of Chinese Academy of Sciences, Beijing 100049, China.}
\affiliation{Departament de F\'isica de la Mat\`eria Condensada, Universitat de Barcelona, Mart\'i i Franqu\`es 1, E08028 Barcelona, Spain}
\author{Abhik Basu}\email{abhik.123@gmail.com, abhik.basu@saha.ac.in}
\affiliation{Theoretical Physics Division, Saha Institute of
Nuclear Physics, 1/AF Bidhannagar, Calcutta 700064, West Bengal, India}
\date{\today}

\begin{abstract}
We investigate pattern formation in generic two-species reaction-diffusion systems with mutual chemotaxis. We show that tuning the chemotaxis strength can give rise to striking nonmonotonic variation of patterns, opening the possibility of re-entrant transitions between patterned and spatially uniform steady states. In certain regimes, chemotaxis also induces additional heretofore unexplored instabilities. Furthermore, varying the strength of chemotaxis can drive morphological transitions between spot and stripe patterns.

    
\end{abstract}

\maketitle


Pattern formation is a hallmark of nonequilibrium systems.  Turing demonstrated that diffusion can destabilize a homogeneous steady state and generate spatial patterns in reaction--diffusion systems when the diffusivities of the interacting species differ sufficiently~\cite{thermohaline_convec, bin_liq_convec, strogatz, turing}. Since then, reaction--diffusion models have become a paradigmatic framework for studying pattern formation, bifurcations, and the competition between multiple instabilities in spatially extended systems~\cite{ricard, predator, *predator1, post_turing, post_turing_2, rmp, book}. Canonical examples, e.g., the Selkov and Brusselator models, exhibit rich phase behavior arising from the interplay of homogeneous and finite-wavelength (Turing) instabilities, including codimension-two bifurcation points~\cite{bruss, bruss1, bruss2, ab_jkb}.

Chemotaxis provides a mechanism for spatial organization by coupling particle motion to chemical gradients, which is fundamentally different from biochemical reactions. It is central to collective migration in biological systems, such as bacterial colonies, tissue development, and immune responses, and also plays an important role in synthetic active matter, phoretic colloids, and enzyme-mediated transport~\cite{lowen, bact_colony, immune, synthetic, recons, artificial, cross_ensyme, frey_enzyme}. Keller-Segel -type models and their extensions have revealed a broad range of chemotactic patterns~\cite{keller_segel, chemo1, chemo2, erwin_chemo_2}. However, despite extensive studies of reaction-diffusion models and chemotaxis separately, their combined effect on pattern selection and universal features, if any, remains poorly understood.

Here we investigate Turing instabilities in a  generic coupled two-species spatially-extended reaction-diffusion system that incorporates both diffusion and mutual chemotaxis between the species. Our studies reveal complex dependence, including nonmonotonic and re-entrant behavior,  of the pattern on the chemotaxis parameters. For explicit calculations, we study a chemotactic extension of the spatially extended version of the Selkov model for glycolysis in two dimensions (2D). Combining linear stability analysis with extensive direct numerical simulations (DNS), we show that chemotaxis qualitatively modifies the instability landscape and the resulting patterns. In particular, the selected wavevector $k_c$ exhibits a pronounced dependence, which can even be nonmonotonic, on the chemotactic couplings. In the DNS studies, this dependence manifests itself as clear variations in the mean pattern amplitudes with  chemotaxis strengths. Our DNS studies also reveal chemotaxis-driven transitions between spot and stripe patterns. Our results demonstrate that chemotaxis provides a powerful mechanism for tuning patterns in reaction--diffusion models, uncovering a new route to controlling nonequilibrium structures in chemically active systems.

We now set up our model and use it to derive the above results. The equations of motion for a generic two-species reaction-diffusion model with species densities $\phi$ and $\psi$ with onsite reactions and mutual chemotaxis have the form
    \begin{eqnarray}
    \frac{\partial \phi}{\partial t} &=&f_\phi(\phi,\psi) + D_1\nabla^2 \phi+\xi_1 {\boldsymbol\nabla}\cdot (\phi{\boldsymbol\nabla}\psi),\label{modbasica}\\
 \frac{\partial \psi}{\partial t} &=& f_\psi(\phi,\psi) + D_2\nabla^2 \psi+\xi_2 {\boldsymbol\nabla}\cdot (\psi{\boldsymbol\nabla}\phi),\label{modbasicb}
\end{eqnarray}
 where $f_\phi(\phi,\psi),\,f_\psi(\phi,\psi)$ are {\it nonconserving} reaction functions that model local (onsite) reactions between the two species, and $D_1, D_2>0$ are the two diffusivities,  and $\xi_1,\xi_2$ are the chemotaxis couplings, which can be individually positive or negative. Physically, when $\xi_1\xi_2>0$, the particles mutually attract (if $\xi_1,\xi_2>0$) or repel (if $\xi_1,\xi_2<0$), a behavior we name reciprocal chemotaxis. In contrast, when $\xi_1\xi_2<0$, one species is attracted towards the other, while the latter is repelled, resulting in chasing (pursuer--evader) dynamics, which we call nonreciprocal chemotaxis. 
 At the fixed points corresponding to the uniform states
 \begin{eqnarray}
     f_\phi(\phi_0,\psi_0)=0=f_\psi(\phi_0,\psi_0).
 \end{eqnarray}
 Expanding (\ref{modbasica}) and (\ref{modbasicb}) around the uniform state with $\phi=\phi_0+u$, $\psi=\psi_0+v$, we obtain  to the linear order in $u$ and $v$ 
 \begin{eqnarray}
     \partial_t u &=& D_1\nabla^2 u + F_{11}u + F_{12}v+ \xi_1\phi_0 \nabla^2 v, \label{geneq1}\\
     \partial_t v &=& D_2 \nabla^2 v + F_{21}u + F_{22} v + \xi_2 \psi_0\nabla^2 u, \label{geneq2}
 \end{eqnarray}
 $F_{11}=\partial f_\phi/\partial\phi,F_{12}=\partial f_\phi/\partial\psi, F_{21}=\partial f_\psi/\partial\phi, F_{22}=\partial f_\psi/\partial\psi$, 
 with all these derivatives evaluated at $\phi=\phi_0,\psi=\psi_0$. Now, assuming time-dependence of the form $\exp(\Lambda t)$, where $\Lambda$ is a growth rate, i.e., setting $u({\bf x}, t), v({\bf x}, t)\sim \exp(\Lambda t)$, we get for the linear stability matrix $J$ for the pair of equations (\ref{geneq1}) and (\ref{geneq2})  in the Fourier space
 \begin{equation}
 J(k^2) = \left(\begin{array}{cc}
                 F_{11} - D_1k^2 & F_{12}-\xi_1\psi_0k^2\\
                 F_{21} -\xi_2\psi_0 k^2 & F_{22} - D_2k^2
                \end{array}
\right).\label{j-matrix}
\end{equation}
The corresponding eigenvalues, labeled by wavevector $k$, are given by
\begin{eqnarray}
 \Lambda_\pm (k^2) &=&\frac{1}{2}[{\tt Tr}\pm \sqrt{{\tt Tr}^2 - 4\,\Delta}],\label{eigen1}
\end{eqnarray}
where ${\tt Tr}$ and $\Delta$, respectively, are the trace and determinant of  $J(k^2)$ in (\ref{j-matrix}) above. Straightforward calculations give
\begin{eqnarray}
  &&\Lambda_\pm (k^2) = \frac{1}{2}\bigg[F_{11}+F_{22}-(D_1+D_2)k^2 \pm \bigg\{[F_{11}+F_{22}\nonumber \\&&-(D_1+D_2)k^2]^{1/2}-4[F_{11}F_{22}-F_{12}F_{21}-k^2(D_2F_{11}+D_1F_{22})\nonumber \\&& +k^2(F_{21}\xi_1\phi_0+F_{12}\xi_2\psi_0)+(D_1D_2 -\xi_1\xi_2 \phi_0\psi_0)k^4]\bigg\}^{1/2}\bigg].  \label{full-Lambda}
\end{eqnarray}
 At $k=0$, ${\tt Tr}(k^2=0)=F_{11}+F_{22}=0$ gives the instability threshold with  frequency $\omega_0=\pm \sqrt{\Delta(k=0)}=\pm\sqrt{F_{11}F_{22}-F_{12}F_{21}}$. This is the zero-wvevector Hopf bifurcation between a spatially uniform state and an oscillatory state with a frequency $\omega_0$. For ${\tt Tr}(k^2)<0$, consider the finite wavevector, zero frequency saddle-node instability, which is our focus. 
To determine $k_c$, the selected wavevector for the instability, we argue that it is the wavevector $k$ for which $\Delta(k^2)$ is a minimum~\cite{ab_jkb}. Hence, we set
\begin{equation}
    \frac{\partial \Delta(k^2)}{\partial k^2}|_{k^2=k_c^2} =0,\;
    \frac{\partial}{\partial k^2}\frac{\partial \Delta(k^2)}{\partial k^2}\bigg|_{k^2=k_c^2}>0,\label{inst-gen}
\end{equation}
together with $\Delta(k_c^2)=0$ at the instability threshold. Solving,
\begin{equation}
    k_c^2 =\frac{D_2F_{11}+D_2F_{22}-(F_{21}\xi_1\phi_0+F_{12}\xi_2\psi_0)}{2(D_1D_2 -\xi_1\xi_2 \phi_0\psi_0)},\label{kc2-gen}
\end{equation}
together with 
   $D_1 D_2 -\xi_1\xi_2 \phi_0\psi_0>0$. 
 Notice that in (\ref{kc2-gen}), the numerator depends {\it linearly} on $\xi_1,\xi_2$, but the denominator depends on them {\it bilinearly}. This can potentially introduce nonmonotonic dependence of $k_c^2$ on $\xi_1,\xi_2$, provided  $F_{21}\xi_1\phi_0+F_{12}\xi_2\psi_0>0$. Furthermore, as $F_{21}\xi_1\phi_0+F_{12}\xi_2\psi_0$ turns negative, possible if $\xi_1\xi_2>0$ and sufficiently large, $k^2=k_c^2 $ gives the {\it maximum} of $\Delta(k^2)$, signaling a chemotaxis-induced instability; see associated long paper (ALP)~\cite{alp}. Whether or not the above-mentioned nonmonotonic behavior or the chemotaxis-induced instability will be displayed depends on the specific model under consideration. To study this more explicitly and investigate the nonlinear effects, we study the generalized Selkov model for glykolysis that includes chemotaxis between the two species of densities $\phi$ and $\psi$. The dynamical equations read 
\begin{eqnarray}
 \frac{\partial \phi}{\partial t} &=&-\phi + a\psi + \phi^2\psi + D_1\nabla^2 \phi +\xi_1 {\boldsymbol\nabla}\cdot (\phi{\boldsymbol\nabla}\psi),\label{mod1}\\
 \frac{\partial \psi}{\partial t} &=&  b - a\psi -\phi^2 \psi + D_2\nabla^2 \psi+\xi_2 {\boldsymbol\nabla}\cdot (\psi{\boldsymbol\nabla}\phi).\label{mod2}
\end{eqnarray}
Here, $a>0,b>0$ are model parameters that control the reactions. In this model, $\phi_0=b,\psi_0=b/(a+b^2)$~\cite{ab_jkb}. By applying the logic of calculations outlined above for the general case, we can calculate $k_c^2$ from (\ref{mod1}) and (\ref{mod2}) in a linear stability analysis that also shows complex dependence of $k_c^2$ on $\xi_1,\xi_2$; see Ref.~\cite{alp}. 
For simplicity and to reduce the number of parameters in the DNS studies to complement our linear stability analysis results, we focus on two extreme cases, {\it viz.}  $\xi_1=\xi_2=\xi$ (fully reciprocal chemotaxis), or $\xi_1=-\xi_2=\xi)$ (extreme nonreciprocal chemotaxis), where $\xi$ itself can be positive and negative. In this case, the chemotaxis-induced instability occurs for
\begin{equation}
    D_1D_2<\xi^2\frac{b^2}{a+b^2}.\label{chemo-induced-inst}
\end{equation}
Thus, this instability can be found only with reciprocal chemotaxis for sufficiently strong $\xi^2$; it has no analog in the nonreciprocal case.
We explore  the complex dependence of $k_c^2$ on the single chemotaxis parameter $\xi$. Considering $\xi_1=\xi=\xi_2$,
\begin{equation}
    k_c^2=\frac{-D_1(a+b^2)+D_2\frac{b^2-a}{a+b^2}+\xi b \frac{b^2-a}{a+b^2}}{2[D_1D_2-\frac{\xi^2b^2}{a+b^2}]},\label{kc-xi-DNS-reci}
\end{equation}
in the fully reciprocal chemotaxis case, and with $\xi_1=\xi=-\xi_2$
\begin{equation}
    k_c^2=\frac{-D_1(a+b^2)+D_2\frac{b^2-a}{a+b^2}+\xi b \frac{a+3b^2}{a+b^2}}{2[D_1D_2+\frac{\xi^2b^2}{a+b^2}]}\label{kc-xi-DNS-nonreci}
\end{equation}
in the extreme nonreciprocal chemotaxis case. 
Consider first Eq.~(\ref{kc-xi-DNS-reci}) for the reciprocal case. For $\xi>0$ if $b^2<a$, the numerator of (\ref{kc-xi-DNS-reci}) that depends {\it linearly} on $\xi>0$, {\it decreases} as $\xi$ rises from zero, which has the effect of {\it decreasing} $k_c^2$. The denominator in (\ref{kc-xi-DNS-reci}) that depends {\it quadratically} on $\xi$ also {\it decreases}, which however tends to {\it increase} $k_c$. Therefore, for sufficiently small $\xi$, $k_c^2$ should {\it decrease} with increasing $\xi$, but for sufficiently large $\xi>\xi_c|_\text{ reci}$, a finite crossover threshold, $k_c^2$ should start to rise. Thus $k_c^2$ should have a minimum at $\xi=\xi_c|_\text{ reci}>0$. By setting $\partial k_c^2/\partial\xi=0$ and using (\ref{kc-xi-DNS-reci}), we get
\begin{eqnarray}
 &&\xi_c|_\text{ reci}=   \frac{a+b^2}{b(b^2-a)}\bigg[D_1(a+b^2)-D_2\frac{b^2-a}{a+b^2}\nonumber \\&& + \{[D_1 (a+b^2) - D_2\frac{b^2-a}{a+b^2}]^2 -D_1D_2 \frac{(b^2-a)^2}{a+b^2}\}^{1/2}\bigg].\nonumber \\
\end{eqnarray}
This holds for $\xi>0$ with $b^2<a$ and $\xi<0$ for $b^2>a$. For real $\xi_c$, we must have the discriminant $\delta_\text{reci}>0$, giving $k_c^2|_\text{reci}$:
\begin{eqnarray}
  &&  \delta_\text{reci}\equiv \bigg[D_1 (a+b^2) - D_2\frac{b^2-a}{a+b^2}\bigg]^2 -D_1D_2 \frac{(b^2-a)^2}{a+b^2}>0,\nonumber \\
 &&   k_c^2|_\text{reci} = \frac{\sqrt{\delta_\text{reci}}}{2\bigg[D_1D_2 -\frac{\xi_c^2|_\text{reci} b^2}{a+b^2}\bigg]},\label{kc-reci}
\end{eqnarray}
 the minimum of $k_c^2$ for $\xi>0$ when $b^2<a$, or for $\xi<0$ when $b^2>a$. Now consider a system of linear size $L$, such that $Lk_c|_\text{reci}\ll 1$, or equivalently $L/\lambda_c|_\text{reci}\ll 1$, where the {\it maximum} wavelength $\lambda_c|_\text{reci}\equiv 2\pi/k_c|_\text{reci}$.  In this case, the system should appear to be effectively uniform. Thus,  for  $L\ll \lambda_c|_\text{reci}$, the system can make a transition from a patterned state for low $\xi$ to a re-entrant patterned state for high $\xi$ via a uniform state at intermediate $\xi\approx \xi_c|_\text{reci}$, as $\xi<0$ is tuned from 0 to high negative through $\xi_c|_\text{reci}<0$.For $\xi<0$ when $b^2<a$, or for $\xi>0$ when $b^2>a$, $k_c^2$ rises monotonically with $|\xi|$. See Fig.~\ref{fig:fig_R_Xineg}(a) and Fig.~\ref{fig:fig_R_Xipos}(a).

We now consider the corresponding extreme nonreciprocal case and study the $\xi$-dependence of (\ref{kc-xi-DNS-nonreci}). For $\xi>0$, the numerator of (\ref{kc-xi-DNS-nonreci}) {\it grows} linearly with $\xi$, which {\it enhances} $k_c^2$. However, the denominator of (\ref{kc-xi-DNS-nonreci})  {\it grows} quadratically with $\xi$, which {\it reduces} $k_c^2$. This suggests a maximum of $k_c^2$ at some finite $\xi=\xi_c|_\text{ nonreci}$. 
Setting $\partial k_c^2/\partial\xi=0$, we find
\begin{eqnarray}
    &&\xi_c|_\text{nonreci}=\frac{a+b^2}{b(a+3b^2}\bigg[D_1(a+b^2)-D_2\frac{b^2-a}{a+b^2}\nonumber \\&&+\{[D_1(a+b^2)-D_2\frac{b^2-a}{a+b^2}]^2+ (a+3b^2)\frac{D_1D_2b^2}{a+b^2}\}^{1/2}\bigg],\nonumber \\
 &&  k_c^2|_\text{nonreci}=\frac{\sqrt{\delta_\text{nonreci}}}{2[D_1D_2+\frac{\xi_\text{nonreci}^2b^2}{a+b^2}]},\label{kc-nonreci}\\
   && \delta|_\text{nonreci}=D_1(a+b^2)-D_2\frac{b^2-a}{a+b^2}]^2+ (a+3b^2)\frac{D_1D_2b^2}{a+b^2},\nonumber 
 \end{eqnarray}
which gives the maximum of $k_c^2$ for $\xi>0$ for both $b^2>a$ and $b^2<a$. 
Consider a system of linear finite size $L >\lambda_c|_\text{nonreci}$, where $\lambda_c|_\text{nonreci}\equiv 2\pi/k_c|_\text{nonreci}$, but $L\ll 2\pi k_c^{-1}(\xi=0)$ or $L\ll 2\pi k_c^{-1}\xi\rightarrow \infty$. In this case, for very small of large $\xi$, the system may appear uniform, but for intermediate values of $\xi>0$, in the neighborhood of $\xi_c|_\text{nonreci}$ a pattern should form, giving a re-entrant transition from uniform state with small $\xi$ to a patterned state for $\xi$ around $\xi_c|_\text{nonreci}$, and ultimately to a re-entrant uniform state for large $\xi>0$. Although nonmonotonic dependence of $k_c^2$ on $\xi$ is possible for specific fully reciprocal or extreme nonreciprocal cases, the nature of the nonmonotonic dependence is very different in the two cases. For $\xi<0$, $k_c^2$ decreases monotonically as $\xi$ becomes large negative. See Fig.~\ref{fig:fig_NR_Xipos}(a) and Fig.~\ref{fig:fig_NR_Xineg}(a).

To complement the above results, we numerically solve Eqs.~(\ref{mod1}) and (\ref{mod2}) by using pseudo-spectral methods with $\xi_1=\xi=\pm\xi_2$. We set $a= 0.015, b = 10$, $D_1=0.01$ with $D=1000$ for which our DNS studies does not show any sustained oscillations, and use square geometry with a linear system size $L=128$. 
To quantify how the patterns depend on $\xi_1,\xi_2$, we define $\Delta_\phi$ and $\Delta_\psi$ as
\begin{eqnarray}
   && \Delta_a\equiv \bigg[{\sum_{x,y}[a(x,y)-\overline a]^2\frac{1}{L^2}}\bigg]^{1/2},\label{Del-phi-a}
\end{eqnarray}
to be measured in the steady states,
where $a=\phi,\psi$, $\overline\phi$ and $\overline\psi$ are the spatial mean values of $\phi(x,y)$ and $\psi(x,y)$, respectively, in the steady states, and hence time-independent. Quantities $\Delta_\phi$ and $\Delta_\psi$ are, in fact, measures of spatial averages of the pattern amplitudes (see below): In the uniform states, $\Delta_\phi=0=\Delta_\psi$, just like pattern amplitudes themselves. However, in the patterned 
states, $\Delta_\phi>0,\,\Delta_\psi>0$. The variation (including nonmonotonic behavior) of $k_c^2$ with $\xi_1,\xi_2$ in the linear stability analysis may get reflected in the DNS studies results in the variation of the periodicity and amplitudes or both of the pattern with $\xi_1,\xi_2$. See Fig.~\ref{fig:fig_R_Xineg}(b),(c),(d),(e), Fig.~\ref{fig:fig_R_Xipos}(b),(c),(d),(e), Fig.~\ref{fig:fig_NR_Xipos}(b),(c),(d),(e) and Fig.~\ref{fig:fig_NR_Xineg}(b),(c),(d),(e) for pattern snapshots the variation of $\Delta_a$ with $\xi$ in  different reciprocal and nonreciprocal cases. 

\begin{widetext}

\begin{figure}[htb]
\includegraphics[width=\columnwidth]{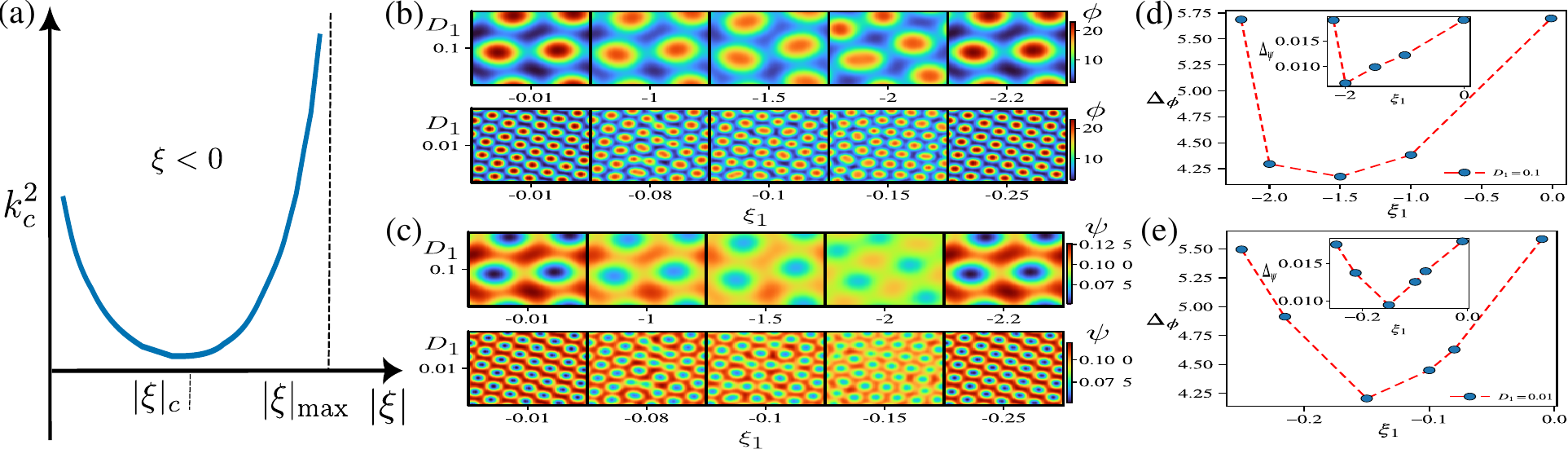}
\caption{Instabilities and patterns for $\xi_1=\xi=\xi_2<0$. 
(a) $k_c^2$ as a function of $\xi<0$, showing a {\it nonmonotonic} behavior. (b) and (c) Snapshots showing steady patterns for $D=1000$ and $D_1=0.01 \text{(b)}, 0.1 \text{(c)} $, showing {\it nonmonotonic} variation of the pattern amplitudes as $\xi$  varies. (d) and (e) Plots of $\Sigma_\phi$ and $\Sigma_\psi$ versus $\xi$ corresponding to the snapshots in (b) and (c) respectively. See text.}
\label{fig:fig_R_Xineg}
\end{figure}

\begin{figure}[htb]
\includegraphics[width=\columnwidth]{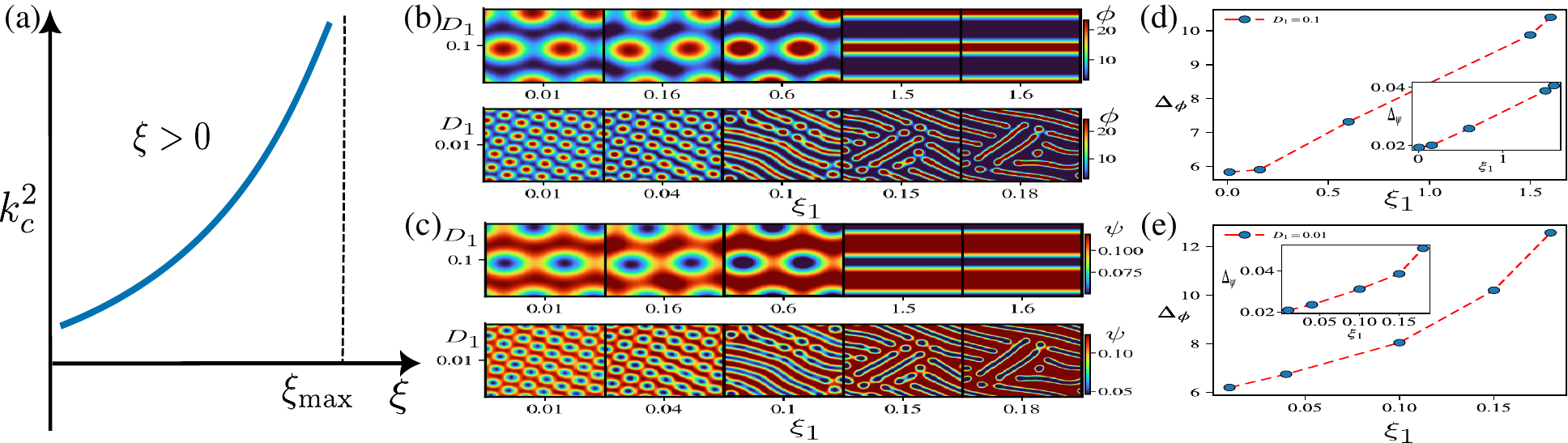}
\caption{Instabilities and patterns for $\xi_1=\xi=\xi_2>0$. 
(a) $k_c^2$ as a function of $\xi>0$, showing a {\it monotonic} behavior. (b) and (c) Snapshots showing steady patterns for $D=1000$ and $D_1=0.01 \text{(b)}, 0.1 \text{(c)} $, showing {\it monotonic} variation of the pattern amplitudes as $\xi$  varies. (d) and (e) Plots of $\Sigma_\phi$ and $\Sigma_\psi$ versus $\xi$ corresponding to the snapshots in (b) and (c) respectively. See text.}
\label{fig:fig_R_Xipos}
\end{figure}

\begin{figure}[htb]
\includegraphics[width=\linewidth]{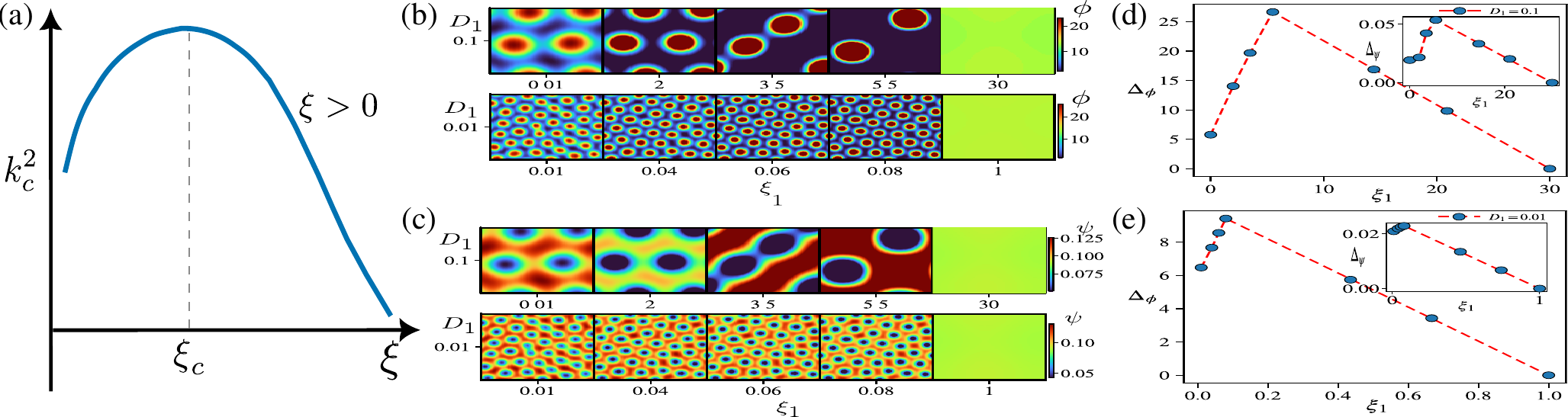}
\caption{Instabilities and patterns for $\xi_1=\xi=-\xi_2>0$.
(a) $k_c^2$ as a function of $\xi>0$, showing a {\it nonmonotonic} behavior. (b) and (c) Snapshots showing steady patterns for $D=1000$ and $D_1=0.01 \text{(b)}, 0.1 \text{(c)} $, showing {\it nonmonotonic} variation of the pattern amplitudes as $\xi$  varies. (d) and (e) Plots of $\Sigma_\phi$ and $\Sigma_\psi$ versus $\xi$ corresponding to the snapshots in (b) and (c) respectively. See text.}
\label{fig:fig_NR_Xipos}
\end{figure}

\begin{figure}[htb]
\includegraphics[width=\columnwidth]{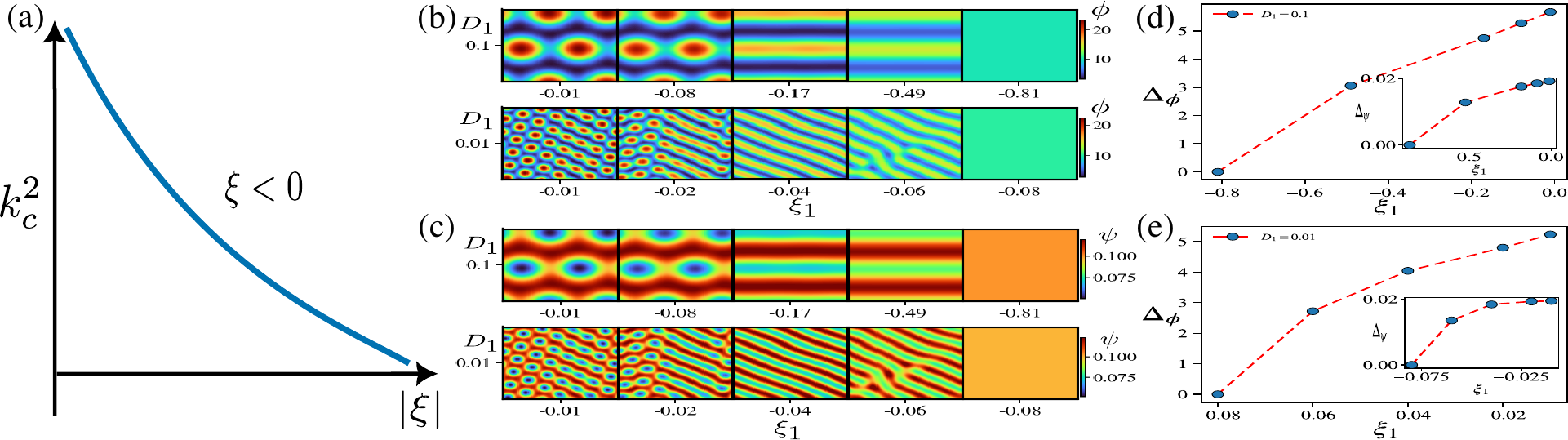}
\caption{Instabilities and patterns for $\xi_1=\xi=-\xi_2<0$.
(a) $k_c^2$ as a function of $\xi<0$, showing a {\it monotonic} behavior. (b) and (c) Snapshots showing steady patterns for $D=1000$ and $D_1=0.01 \text{(b)}, 0.1 \text{(c)} $, showing {\it monotonic} variation of the pattern amplitudes as $\xi$  varies. (d) and (e) Plots of $\Sigma_\phi$ and $\Sigma_\psi$ versus $\xi$ corresponding to the snapshots in (b) and (c) respectively. See text.}
\label{fig:fig_NR_Xineg}
\end{figure}

\end{widetext}
All of Fig.~\ref{fig:fig_R_Xineg}, Fig.~\ref{fig:fig_R_Xipos}, Fig.~\ref{fig:fig_NR_Xipos} and Fig.~\ref{fig:fig_NR_Xineg} reveal a striking one-to-one correspondence between the linear stability results on the dependence of $k_c^2$  and the mean amplitudes of the pattern snapshots given by $\Sigma_\phi,\Sigma_\psi$ on $\xi$. Surprisingly, we also find spot-to-stripe transitions of the patterns in  Fig.~\ref{fig:fig_R_Xipos} and Fig.~\ref{fig:fig_NR_Xineg} as $\xi$ is varied, in spite of the rotational invariance of the governing dynamical equations. Stripe patterns are found to be linearly stable, as shown in ALP~\cite{alp}. However, this stability condition is identical to the stability condition (\ref{chemo-induced-inst}) of the pattern itself. This strongly suggests that the spot-to-stripe transition should be a nonlinear phenomenon, as discussed in  details in ALP~\cite{alp}.

To summarize, we have investigated Turing instabilities in generic two-component nonconserved reaction--diffusion systems generalized to include mutual chemotaxis. Linear stability analysis reveals that the selected wavevector $k_c$ exhibits a sensitive dependence on the chemotaxis couplings, including nonmonotonic behavior whose form depends on the underlying chemotaxis. To quantitatively study these effects, we studied a spatially extended Selkov model for glycolysis with chemotaxis in both fully reciprocal and extreme nonreciprocal cases, parameterized by $\xi$.
Linear stability analysis reveals a strong dependence, including a nonmonotonic dependence, of $k_c$ on $\xi$, with contrasting natures in the fully reciprocal and extreme nonreciprocal cases. In DNS studies, these dependences manifest themselves in the variation of the pattern amplitudes with $\xi$, once again revealing the corresponding contrasting behaviors in the fully reciprocal and extreme nonreciprocal cases, which establishes a one-to-one correspondence between the linear stability and DNS results. Our linear stability analysis predictions of a novel chemotaxis-induced instability for reciprocal chemotaxis with any sign of $\xi$ is corroborated by our DNS results. 
The generality of our results strongly suggests that these can be observed in any general two-species reaction diffusion model with mutual chemotaxis. We expect our studies will provide impetus to experimental realizations in relevant biological and artificial in-vitro systems.

{\it Acknowledgment:}
A.B. thanks Alexander von Humboldt Stiftung (Germany) for partial financial support through their research group linkage programme (2024) and ANRF (India) for partial financial support through the ARG (MATRICS) programme (file no.: ANRF/ARGM/2025/000461/TS).
 

\bibliography{ref_chemo_short-1}
\end{document}